\documentclass[aps,prl,twocolumn,superscriptaddress,showpacs]{revtex4}

\usepackage{graphicx}% Include figure files
\usepackage{dcolumn}% Align table columns on decimal point
\usepackage{bm}% bold math
\usepackage{setspace}
\usepackage{epsfig}
\usepackage{epstopdf}
\def\met{$E\!\!\!/_T$}
\def\et{$E_T$}
\def\pt{$p_T$}
\def\etjet{$E_{T}^{jet}$}
\def\wmt{$m_T^W$}

\def\nth{$n^{th}$}

\def\wjets{$W+{\rm jet(s)}$ }

\def\vjets{$V+{\rm jet(s)}$}
\def\wnjet{$W+n$-${\rm jet}$ }

\def\wonejet{$W+1$-${\rm jet}$ }

\def\wgenjet{$W+\geq n$-${\rm jet}$ }

\def\wgeonejet{$W+\geq 1$-${\rm jet}$}
\def\wge2jet{$W+\geq 2$-${\rm jet}$}
\def\wevgenjet{$W\to e\nu +\geq n$-${\rm jet}$}

\def\wev{$W\to e\nu$ }

\def\diffsigmaBR{$d\sigma(p\bar{p}\rightarrow W+\ge~n$-${\rm jet})/dE_{T}^{nth-jet} \times {\cal B}(W \rightarrow e\nu)$}
\def\diffsigma{$d\sigma(W\to e\nu+\ge~n$-${\rm jet})/dE_{T}^{nth-jet}$}
\def\totalnsigmaBR{$\sigma_n = \sigma(p\bar{p}\rightarrow W+\geq n$-${\rm jet}; E_T^{nth-jet}>25~{\rm GeV}) \times {\cal B}(W \rightarrow e\nu)$}
\def\totalsigmaBR{$\sigma(p\bar{p}\rightarrow W+\geq n$-${\rm jet}; E_T^{nth-jet}>25~{\rm GeV}) \times {\cal B}(W \rightarrow e\nu)$}
\def\totalnsigma{$\sigma_n = \sigma(W\to e\nu+\geq n$-${\rm jet}; E_T^{nth-jet}>25~{\rm GeV})$}

\def\be{\begin{equation}}
\def\ee{\end{equation}}

\begin{document}
\title{Measurement of the cross section for $W$-boson production in association with jets in $p\bar{p}$ collisions at $\sqrt{s}=1.96$ TeV}
\affiliation{Institute of Physics, Academia Sinica, Taipei, Taiwan 11529, Republic of China} 
\affiliation{Argonne National Laboratory, Argonne, Illinois 60439} 
\affiliation{Institut de Fisica d'Altes Energies, Universitat Autonoma de Barcelona, E-08193, Bellaterra (Barcelona), Spain} 
\affiliation{Baylor University, Waco, Texas  76798} 
\affiliation{Istituto Nazionale di Fisica Nucleare, University of Bologna, I-40127 Bologna, Italy} 
\affiliation{Brandeis University, Waltham, Massachusetts 02254} 
\affiliation{University of California, Davis, Davis, California  95616} 
\affiliation{University of California, Los Angeles, Los Angeles, California  90024} 
\affiliation{University of California, San Diego, La Jolla, California  92093} 
\affiliation{University of California, Santa Barbara, Santa Barbara, California 93106} 
\affiliation{Instituto de Fisica de Cantabria, CSIC-University of Cantabria, 39005 Santander, Spain} 
\affiliation{Carnegie Mellon University, Pittsburgh, PA  15213} 
\affiliation{Enrico Fermi Institute, University of Chicago, Chicago, Illinois 60637} 
\affiliation{Comenius University, 842 48 Bratislava, Slovakia; Institute of Experimental Physics, 040 01 Kosice, Slovakia} 
\affiliation{Joint Institute for Nuclear Research, RU-141980 Dubna, Russia} 
\affiliation{Duke University, Durham, North Carolina  27708} 
\affiliation{Fermi National Accelerator Laboratory, Batavia, Illinois 60510} 
\affiliation{University of Florida, Gainesville, Florida  32611} 
\affiliation{Laboratori Nazionali di Frascati, Istituto Nazionale di Fisica Nucleare, I-00044 Frascati, Italy} 
\affiliation{University of Geneva, CH-1211 Geneva 4, Switzerland} 
\affiliation{Glasgow University, Glasgow G12 8QQ, United Kingdom} 
\affiliation{Harvard University, Cambridge, Massachusetts 02138} 
\affiliation{Division of High Energy Physics, Department of Physics, University of Helsinki and Helsinki Institute of Physics, FIN-00014, Helsinki, Finland} 
\affiliation{University of Illinois, Urbana, Illinois 61801} 
\affiliation{The Johns Hopkins University, Baltimore, Maryland 21218} 
\affiliation{Institut f\"{u}r Experimentelle Kernphysik, Universit\"{a}t Karlsruhe, 76128 Karlsruhe, Germany} 
\affiliation{Center for High Energy Physics: Kyungpook National University, Daegu 702-701, Korea; Seoul National University, Seoul 151-742, Korea; Sungkyunkwan University, Suwon 440-746, Korea; Korea Institute of Science and Technology Information, Daejeon, 305-806, Korea; Chonnam National University, Gwangju, 500-757, Korea} 
\affiliation{Ernest Orlando Lawrence Berkeley National Laboratory, Berkeley, California 94720} 
\affiliation{University of Liverpool, Liverpool L69 7ZE, United Kingdom} 
\affiliation{University College London, London WC1E 6BT, United Kingdom} 
\affiliation{Centro de Investigaciones Energeticas Medioambientales y Tecnologicas, E-28040 Madrid, Spain} 
\affiliation{Massachusetts Institute of Technology, Cambridge, Massachusetts  02139} 
\affiliation{Institute of Particle Physics: McGill University, Montr\'{e}al, Canada H3A~2T8; and University of Toronto, Toronto, Canada M5S~1A7} 
\affiliation{University of Michigan, Ann Arbor, Michigan 48109} 
\affiliation{Michigan State University, East Lansing, Michigan  48824} 
\affiliation{University of New Mexico, Albuquerque, New Mexico 87131} 
\affiliation{Northwestern University, Evanston, Illinois  60208} 
\affiliation{The Ohio State University, Columbus, Ohio  43210} 
\affiliation{Okayama University, Okayama 700-8530, Japan} 
\affiliation{Osaka City University, Osaka 588, Japan} 
\affiliation{University of Oxford, Oxford OX1 3RH, United Kingdom} 
\affiliation{University of Padova, Istituto Nazionale di Fisica Nucleare, Sezione di Padova-Trento, I-35131 Padova, Italy} 
\affiliation{LPNHE, Universite Pierre et Marie Curie/IN2P3-CNRS, UMR7585, Paris, F-75252 France} 
\affiliation{University of Pennsylvania, Philadelphia, Pennsylvania 19104} 
\affiliation{Istituto Nazionale di Fisica Nucleare Pisa, Universities of Pisa, Siena and Scuola Normale Superiore, I-56127 Pisa, Italy} 
\affiliation{University of Pittsburgh, Pittsburgh, Pennsylvania 15260} 
\affiliation{Purdue University, West Lafayette, Indiana 47907} 
\affiliation{University of Rochester, Rochester, New York 14627} 
\affiliation{The Rockefeller University, New York, New York 10021} 
\affiliation{Istituto Nazionale di Fisica Nucleare, Sezione di Roma 1, University of Rome ``La Sapienza," I-00185 Roma, Italy} 
\affiliation{Rutgers University, Piscataway, New Jersey 08855} 
\affiliation{Texas A\&M University, College Station, Texas 77843} 
\affiliation{Istituto Nazionale di Fisica Nucleare, University of Trieste/\ Udine, Italy} 
\affiliation{University of Tsukuba, Tsukuba, Ibaraki 305, Japan} 
\affiliation{Tufts University, Medford, Massachusetts 02155} 
\affiliation{Waseda University, Tokyo 169, Japan} 
\affiliation{Wayne State University, Detroit, Michigan  48201} 
\affiliation{University of Wisconsin, Madison, Wisconsin 53706} 
\affiliation{Yale University, New Haven, Connecticut 06520} 
\author{T.~Aaltonen}
\affiliation{Division of High Energy Physics, Department of Physics, University of Helsinki and Helsinki Institute of Physics, FIN-00014, Helsinki, Finland}
\author{J.~Adelman}
\affiliation{Enrico Fermi Institute, University of Chicago, Chicago, Illinois 60637}
\author{T.~Akimoto}
\affiliation{University of Tsukuba, Tsukuba, Ibaraki 305, Japan}
\author{M.G.~Albrow}
\affiliation{Fermi National Accelerator Laboratory, Batavia, Illinois 60510}
\author{B.~\'{A}lvarez~Gonz\'{a}lez}
\affiliation{Instituto de Fisica de Cantabria, CSIC-University of Cantabria, 39005 Santander, Spain}
\author{S.~Amerio}
\affiliation{University of Padova, Istituto Nazionale di Fisica Nucleare, Sezione di Padova-Trento, I-35131 Padova, Italy}
\author{D.~Amidei}
\affiliation{University of Michigan, Ann Arbor, Michigan 48109}
\author{A.~Anastassov}
\affiliation{Rutgers University, Piscataway, New Jersey 08855}
\author{A.~Annovi}
\affiliation{Laboratori Nazionali di Frascati, Istituto Nazionale di Fisica Nucleare, I-00044 Frascati, Italy}
\author{J.~Antos}
\affiliation{Comenius University, 842 48 Bratislava, Slovakia; Institute of Experimental Physics, 040 01 Kosice, Slovakia}
% ...
\collaboration{CDF Collaboration\footnote{With visitors from 
$^a$University of Athens, 15784 Athens, Greece, 
$^b$Chinese Academy of Sciences, Beijing 100864, China, 
$^c$University of Bristol, Bristol BS8 1TL, United Kingdom, 
$^d$University Libre de Bruxelles, B-1050 Brussels, Belgium, 
$^e$University of California Irvine, Irvine, CA  92697, 
$^f$University of California Santa Cruz, Santa Cruz, CA  95064, 
$^g$Cornell University, Ithaca, NY  14853, 
$^h$University of Cyprus, Nicosia CY-1678, Cyprus, 
$^i$University College Dublin, Dublin 4, Ireland, 
$^j$University of Edinburgh, Edinburgh EH9 3JZ, United Kingdom, 
$^k$University of Heidelberg, D-69120 Heidelberg, Germany, 
$^l$Universidad Iberoamericana, Mexico D.F., Mexico, 
$^m$University of Manchester, Manchester M13 9PL, England, 
$^n$Nagasaki Institute of Applied Science, Nagasaki, Japan, 
$^o$University de Oviedo, E-33007 Oviedo, Spain, 
$^p$Queen Mary, University of London, London, E1 4NS, England, 
$^q$Texas Tech University, Lubbock, TX  79409, 
$^r$IFIC(CSIC-Universitat de Valencia), 46071 Valencia, Spain, 
}}
\noaffiliation

\date{\today}
% 
%%% ABSTRACT %%%
%
\begin{abstract}
We present a measurement of the cross section for $W$-boson production in association with jets
in $p\bar{p}$ collisions at $\sqrt{s}=1.96$~TeV. The analysis uses a data sample corresponding to an
integrated luminosity of $320$~${\rm pb^{-1}}$ collected with the CDF II detector. $W$ bosons are identified
in their electron decay channel and jets are reconstructed using a cone algorithm.
For each \wgenjet sample ($n= 1 - 4$) we measure the differential cross section 
\diffsigmaBR~with respect to the transverse energy $E_T$ of the $n^{th}$-highest $E_{T}$ 
jet above $20~{\rm GeV}$, and the total cross section \totalsigmaBR, for a 
restricted $W \rightarrow e\nu$ decay phase space. The cross sections, corrected
for all detector effects, can be directly compared to particle level $W+{\rm jet(s)}$ predictions. We present here
comparisons to leading order and next-to-leading order predictions.
\end{abstract}
% insert suggested PACS numbers in braces on next line
\maketitle
% body of paper here
\vskip -1cm
%
%%% INTRODUCTION %%%
%
Final states containing a vector boson $V$ ($V=W,Z$) and multiple jets
(\vjets) are a key signal channel for important standard model
(SM) processes such as $t\bar{t}$ or single top production, as well
as a search channel for the Higgs boson and for physics beyond the SM.
The production of $V+$jet(s) via quantum
chromodynamics (QCD) presents a very large background to
these processes. The ability to describe it accurately 
is therefore crucial, as well as being a stringent test of the 
power of perturbative QCD predictions.
Consequently, a precise measurement of the 
cross section for QCD \vjets~production is an important component of the hadron collider 
experimental program.
In this paper, we report a measurement~\cite{ref:thesisBDCthesisAM} of the differential cross sections 
for direct \wevgenjet~production as a function of the transverse energy 
\etjet~\cite{ref:coord}~of the $n^{th}$-leading jet (the highest \et~jet for 
\wgeonejet, the second highest \et~jet for \wge2jet, etc.), for $n= 1 - 4$ and $E_T^{jet} > 20$~${\rm GeV}$. 
We also provide the total cross section \totalnsigma~ for $n= 1 - 4$. 
In order to minimize the dependence of the measurement on the modeling of the $W$ boson production and decay 
kinematics, we quote cross sections defined within a limited $W$ decay phase space:
$E_T^{ele}>20$~${\rm GeV}$,~$|\eta^{ele}|<1.1$,~$E_T^{\nu}>30$~${\rm GeV}$, 
and~$m_T^{W}>20$~$\rm{GeV/c^2}$~\cite{ref:coord}. 
The range of \etjet~extends up to $350$~${\rm GeV}$ in the 
\wonejet sample, a significant increase in the measured phase space compared to 
previous \vjets~measurements~\cite{ref:run1Njetrun11jet,ref:d0zjets}. 
Furthermore, the differential spectra presented here are for the
first time corrected for all detector effects and represent
absolute particle level cross sections \cite{ref:particlelevel} free, within systematic uncertainties, of any
experimental bias. As such, they provide a benchmark which can be directly used
for background estimates and for the validation and
tuning of QCD phenomenological models. At the end of this paper we present, as examples, 
comparisons of our results with some of the available predictions. It is important to note that 
the cross section is not corrected for effects resulting from the interaction between the proton and anti-proton 
remnants (the ``underlying event''). Such a correction would introduce into the measurement a dependence 
on theoretical models of the underlying event.

%
%%% DETECTOR :
%
This analysis uses $320 \pm 18$~${\rm pb^{-1}}$ of data collected
using the CDF II~\cite{ref:CDFb} detector during the Tevatron Run II period. The CDF
II detector is a general-purpose 
detector designed to study $p\bar{p}$ collisions at the
Fermilab Tevatron.
Inside a 1.4 T solenoidal magnetic field, a large open-cell drift
chamber and an eight-layer silicon system provide precise
charged-particle tracking information. Outside the solenoid 
electromagnetic and hadronic sampling calorimeters surround the
tracking volume, allowing for the
measurement of particle energies over the range $|\eta| <
3.6$. Finely segmented detectors located at electromagnetic
shower maximum are used for electron identification.
Forward gas Cerenkov detectors measure
the fraction of bunch crossings that result in an inelastic $p\bar{p}$
collision and thereby determine the instantaneous luminosity delivered
to the experiment.

%
%%% EVENT SELECTION :
%
The selection of \wev events proceeds as follows. An online trigger
system selects events containing an electromagnetic calorimeter
cluster with $E_{T}>18$~${\rm GeV}$ associated with a high
\pt~track. Offline, electron candidates are required to pass standard
identification cuts~\cite{ref:tightsel} and to have $E_T^{ele}>20$~${\rm
GeV}$.
The sample is enriched with events containing a neutrino by requiring
that the missing transverse energy $E\!\!\!/_T$~\cite{ref:coord}, corrected for the jet energy
scale (see below) and the
potential presence of muons~\cite{ref:corrMet}, satisfies $E\!\!\!/_T>30~{\rm GeV}$.  To
 reduce background contamination further, the $W$ transverse mass~\cite{ref:coord} $m_T^W$ is
required to satisfy $m_T^W>20$~${\rm GeV/c^2}$.  In addition, $Z\to
e^{+}e^{-}$ events are rejected by a veto algorithm~\cite{ref:tightsel}. 

%%% EVENT SELECTION - JETS :
The jets in each \wev event are reconstructed using the {\sc jetclu} cone algorithm~\cite{ref:jetalg}
with cone radius $R =\sqrt{\Delta\phi^2+\Delta\eta^2} =
0.4$. Starting from seed locations corresponding to calorimeter towers with  $E_T>1$~${\rm GeV}$,
 all nearby towers with $E_T>0.1$~${\rm GeV}$ are used to search for stable cones. 
To resolve ambiguities with overlapping cones, cones sharing an energy fraction greater than 0.75 are merged into 
a single jet; otherwise the shared towers are assigned to 
the closest jet. 
We apply a jet energy scale (JES)
correction~\cite{ref:jetcorr} such that the measured \etjet~is on
average equal to the summed \et~of the particles within the jet cone that
are the result of the $p\bar{p}\to W+X$ interaction. 
Jets are required to have $E_T^{jet}>20$~${\rm GeV}$ and
$|\eta|<2.0$.

To ensure negligible overlap of electron and jet energy
deposits, events are rejected if any jet lies within $\Delta R=0.52$
of the $W$ decay electron~\cite{ref:run1Njetrun11jet}. Using this jet
definition, we divide the inclusive \wev candidate events into \wgenjet
samples, and in each sample form the \etjet~spectrum of the
\nth~\et-ordered jet.

\begin{figure}
\includegraphics[width=0.46\textwidth]{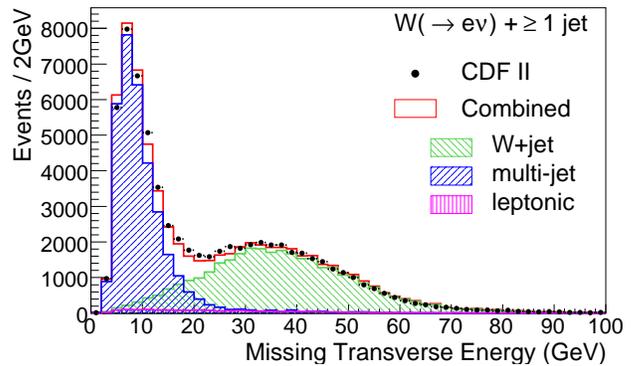}%
\caption{The results of fitting the signal and background \met~template 
distributions to the data in the \wgeonejet~sample before the final \met~cut is applied. 
\label{fig:bkgd}}
\end{figure}

%%% BACKGROUNDS :
%
The processes which contribute background events to our $W$ candidate
sample can be divided into two categories: ``leptonic'' and
``multi-jet''. The leptonic background contains real electrons
and/or neutrinos from boson decay and includes $W\to\tau\nu$, $Z\to
e^+e^-$, $WW$, $W\gamma$, and top pair production. The multi-jet
background arises from QCD interaction events in which one or more
jets are incorrectly reconstructed in the detector as electrons and have
mis-measured energy, resulting in large event \met. Background estimation
proceeds as follows. For each \wgenjet sample, a background-enriched
event sample is constructed by removing the \met~$>30$~${\rm GeV}$
requirement in the $W$ selection. Multi-jet, leptonic, and signal \met~histograms 
are then fit to the data in the range [0,100] GeV. The measured \met~spectrum,
and the result of the fit, are shown in Fig.~\ref{fig:bkgd} for the \wgeonejet~sample. 
For this fit the leptonic background and signal processes are modeled by applying
the $W$ event selection minus the \met~$>30$~${\rm GeV}$ requirement to detector simulated Monte Carlo event
samples of these processes. The multi-jet background is modeled using
an event sample selected from the same $320$~${\rm pb^{-1}}$ analysis
dataset by requiring that at least two of the electron identification
criteria fail. Kinematic cuts are unchanged, resulting in a background-dominated 
sample that accurately reflects the kinematic distributions of the
multi-jet background events in the signal sample. It is necessary to
correct this multi-jet sample for $\sim5\%$ contamination from signal
and leptonic background events, estimated by applying the multi-jet
selection criteria to the Monte Carlo simulations of these processes.
In the fit, only the normalizations of the multi-jet and signal
\met~histograms are allowed to float. The normalization of the
$W\to\tau\nu$ and $Z\to e^+e^-$ histograms relative to the signal is
fixed by the well-established relationships between these cross
sections~\cite{ref:Wlikexsec}. The normalizations of the
$WW$,~$W\gamma$, and top pair production histograms are determined using
the recently measured cross sections for these
processes~\cite{ref:dibosonxsectopxsec}. Once the fit is
performed, the background fractions in each \wgenjet sample are
obtained by integrating the respective histograms, with their fitted
normalization, above the \met~cut of $30$~${\rm GeV}$.  These
fractions are then used to normalize the jet \et~distributions of each
background model relative to the candidates to give the background
correction as a function of \etjet.
This method offers a more accurate description of 
the kinematics of the multi-jet background when compared with previous approaches 
\cite{ref:run1Njetrun11jet}. 
Fig.~\ref{fig:bkgd} demonstrates a successful modeling of the \met~spectrum.
Similarly good
agreement was found in the electron $E_T$ and \wmt~distributions across all jet multiplicities.

%
% PROMOTION :
%
Rarely, reconstructed jets may originate from separate $p\bar{p}$
interactions in the same bunch crossing. 
To account for this effect, corrections to the 
measured cross sections are computed by multiplying the number of
overlapping $p\bar{p}$ interactions, estimated using the primary vertex
multiplicity in the signal sample, with the rate of jet production in
``minimum-bias'' events selected independently of activity in the
central detector. The corrections are less than 2\% in the lowest 
\etjet~bins and decrease rapidly with increasing \etjet.

%%%%%%%%%%%%%%%%%
%
% BKGD SUMMARY & SYSTEMATICS :
%
The total background fraction increases with increasing jet
multiplicity and transverse energy.  At low $E_T^{jet}$ it is 10\%
(40\%) in the $1$-jet ($4$-jet) sample, rising to 90\% at the highest
$E_T^{jet}$ for all jet multiplicities. Multi-jet events contribute $\sim70\%$ of the overall
background in the $1$-jet sample.  At high jet multiplicities and high
$E_T^{jet}$ the contribution from top pair production becomes increasingly
important, climbing to 50\% (80\%) of the total background in the
$2$-jet ($3,4$-jet) sample.
The systematic uncertainty on the background estimate is 15\% at  low $E_T^{jet}$
independent of the jet multiplicity, rising to 50\% (20\%) at the highest 
$E_T^{jet}$ in the $1$-jet ($4$-jet) sample. At low jet multiplicities, this is dominated
by the limited statistics of the multi-jet background sample. At high jet multiplicities, the 12\%
uncertainty on the measured top pair production cross section dominates the total systematic.

%
%%% ACCEPTANCE CORRECTIONS :
%
We have used fully simulated signal Monte Carlo samples to correct the
event yield for the efficiency of the \wev selection criteria.
Samples for each jet multiplicity $n$ were obtained using the {\sc alpgen}
v1.3~\cite{ref:alpgen} event generator for  the $W+n$-parton final state,
and the {\sc pythia} v6.3~\cite{ref:pythia} Monte Carlo program for the parton shower and
hadronization.  {\sc pythia} includes an underlying
event model, hereafter referred to as {\sc tune~a}, which has been tuned to describe Tevatron
data~\cite{ref:TUNEA}.  The efficiency correction factor
is defined as the ratio of two subsets of the generated signal events:
in the numerator, the number of reconstructed events which pass the \wev selection criteria, and in
the denominator, the number of generator-level events which pass the electron, neutrino and
transverse-mass cuts corresponding to our cross section definition.
This is found to be $(60\pm 3)\%$, independent of event jet multiplicity and 
kinematics within the quoted uncertainty.
By comparing $Z\to e^{+}e^{-}$ measured and simulated event samples, we confirm
that electron identification efficiencies are well reproduced by our
Monte Carlo across all measured phase space to within $5\%$.
This uncertainty also covers the observed variation in efficiency obtained
by changing the number of final state partons in the {\sc alpgen}
matrix element (ME)
calculation and the parton showering program from {\sc pythia} to {\sc
herwig} v6.5~\cite{ref:herwig}.  Additionally, the $W$ candidate event
yield must be corrected to account for the efficiency of the online
trigger to accept high \et~electrons. This is independent of jet kinematics and found to be
$96.2\pm0.6\%$~\cite{ref:tightsel}.

%
%%% UNFOLDING CORRECTIONS :
%
A further correction to the event yield is required to form particle level \wevgenjet~cross sections 
as a function of \etjet. This correction factor accounts for the effect of
calorimeter jet energy resolution on the measured cross section and is
determined as follows. Using the {\sc
alpgen-pythia} simulated signal samples, two cross sections in
each \etjet~bin are determined: one defined by clustering generator-level particles
into jets, the other by clustering after detector reconstruction, using the same 
calorimeter level jet definition (including JES correction) as the one used in the data. 
The correction factor for each bin is then defined as the ratio of the particle to calorimeter
level cross section in that bin.  To avoid dependence of such a
correction on the assumed particle level
\etjet~distribution, an iterative procedure is used to reweight the
events at the particle level until the particle level
\etjet~distribution agrees with the corresponding data-unfolded distribution to within the systematic
uncertainties of the measurement. The correction factors vary between $0.95$ and $1.2$ over the measured 
range of \etjet. 
 
%
%%% Systematics related to jet energy measurements:
%
The total systematic uncertainty on the cross section introduced by the jet energy 
measurement ranges between 5\% and 20\%, increasing with increasing~\etjet. This 
is dominated by the approximately 3\%~\cite{ref:jetcorr} uncertainty on the JES correction.
The effect of this uncertainty on the cross section is estimated by applying
this variation to an {\sc alpgen-pythia} simulated signal sample reweighted to
match the data. The sensitivity of the measurement to jet energy resolution
uncertainties, estimated by varying the calorimeter resolution in the simulation,
is much smaller by comparison.

The measured differential cross sections \diffsigma~are listed in Table~\ref{tab:xsecdiff}. 
For each inclusive jet multiplicity sample the cross sections are given with respect to the $E_{T}$ of the
$n^{th}$-leading jet in the sample, $E_{T}^{nth-jet}$. The quoted statistical uncertainties are on the 
event yield in each bin. The systematic uncertainties are the sum in quadrature of the effects 
introduced by the uncertainty on the background estimation, acceptance 
correction, and jet energy measurement. A $5.8\%$ uncertainty in the integrated luminosity
is not included, since this uncertainty is completely correlated between different $E_{T}^{nth-jet}$ bins.
In summary, the total systematic uncertainty on the measured cross
sections is $< 20\%$ at low \etjet~increasing to $50\%-80\%$ at high
\etjet. At low \etjet~this is dominated by
the uncertainty on the jet energy scale, whereas at high \etjet~it is
dominated by the background uncertainty. 
We also provide the total cross section \totalnsigma~for $n= 1-4$: 
$\sigma_1=53.5   \pm 0.6    $(stat.)$ \pm 4.6$(syst.)$ \pm 3.1$(lum.)~pb;
$\sigma_2=6.8   \pm 0.2     $(stat.)$ \pm 1.0$(syst.)$ \pm 0.4$(lum.)~pb;
$\sigma_3=0.84   \pm 0.10   $(stat.)$ \pm 0.21$(syst.)$ \pm 0.05$(lum.)~pb; and 
$\sigma_4=0.074  \pm 0.039  $(stat.)$ \pm 0.035$(syst.)$ \pm 0.004$(lum.)~pb.
The choice of $E_{T}^{nth-jet} > 25~\rm{GeV}$ is made in order to provide a benchmark measurement that is 
less sensitive to the impact of the underlying event, largest at low \etjet~(see below). 
We include for completeness the total inclusive $p\bar{p}\rightarrow W \times {\cal B}(W \rightarrow e\nu)$ 
cross section for the restricted $W \rightarrow e\nu$ decay phase space: 
$\sigma_0=798    \pm 2      $(stat.)$ \pm 40$(syst.)$ \pm 46$(lum.)~pb.

%NEW THEORY COMPARISON PARA
We proceed to compare the measured cross sections to some of the
available theoretical predictions. Leading order (LO) perturbative QCD calculations
exist for the matrix element of $V + n$ partons, with $n \leq 6$.  They are
included in Monte Carlo event
generators~\cite{ref:alpgen,ref:madgraph,ref:gleisberg2003lavesson2005papadopoulos2005} 
where the initial and final state
partons are evolved through a perturbative parton shower (PS) and eventually
hadronized. Additionally, the generator may include a model of the underlying event.
In this LO plus PS approach, ambiguities may arise as a
result of the hard emission of gluon radiation during the parton shower evolution. For
example, a $V+n$-parton event may be reconstructed, after the shower,
with a jet multiplicity $n_j \ne n$, and the question naturally arises as to
whether the event should or should not be counted in the estimate
of the $n_j$-jet cross section. This problem has been studied
extensively in the literature, leading to the development of three 
merging algorithms, usually known as CKKW~\cite{ref:CKKWkrauss2002}, 
Lonnblad's~\cite{ref:lonnblad2002}, and MLM~\cite{ref:mangano2006}. A merging 
algorithm ensures that a given configuration in the multi-jet phase space enters 
into the calculation once and only once.

We present here the 
first comparisons of \wjets implementations of the CKKW and MLM schemes 
to data. We use an implementation of the CKKW scheme hereafter referred to as 
SMPR~\cite{ref:mrenna}.
Details of the generation parameters and systematic uncertainties are
given in~\cite{ref:mrenna} for the SMPR model and
in~\cite{ref:alwall2007} 
for the MLM model. 
The former uses {\sc madgraph} v4~\cite{ref:madgraph} 
for the ME generation, {\sc pythia} v6.3 for the PS, and CTEQ6L1~\cite{ref:pumplin2002} 
parton distribution functions (PDFs), the
latter uses {\sc alpgen} v2.12, {\sc herwig} v6.5, and  CTEQ5L~\cite{ref:lai2000} PDFs respectively.
Following generation, the SMPR and MLM predictions are formed by 
clustering the final 
state particles into jets using the {\sc jetclu} algorithm.
The uncertainties on these predictions cover variation of the
renormalization scale by a factor 0.5--2. This dominates
the overall uncertainty in the absolute rates~\cite{ref:alwall2007}. 

In addition, we present comparisons of our data with next-to-leading order (NLO)
predictions for $W + 1$ and $W + 2$ jets obtained with the {\sc MCFM}~\cite{ref:MCNLO} program. 
These were generated using the CTEQ6.1M PDFs~\cite{ref:pumplin2002} and a
renormalization and factorization scale $\mu = \sqrt{m_W^2+(p_{T}^W)^2}$. 
We define an uncertainty due to the choice of $\mu$ by generating with a lower scale, $\mu = p_T^{jet}$, 
and a higher scale, $\mu = 2*\sqrt{m_W^2+(p_{T}^{W})^2}$. Additionally, the variation due to 
the uncertainty on the PDFs has been computed using the Hessian method~\cite{ref:pumplin2002}. 
This PDF uncertainty is also broadly applicable to the SMPR and MLM predictions.

In the case of the NLO predictions, the final states are 
not evolved through a parton shower nor hadronized. 
Jets are reconstructed with a cone algorithm $R=0.4$, such that two partons are merged 
if they are within $1.3\times R$ of each other and within $R$ of the resulting
jet centroid~\cite{ref:Ellis1993tq}.
This is still considered to be sufficient to give a reasonable 
description of the perturbative structure of the jet~\cite{ref:campbell2006}. 
However, before comparing with data, the
non-perturbative effects of hadronization and the underlying event
have to be considered. We have estimated, using {\sc pythia} {\sc tune~a}, the impact of these two effects. 
The effect of the underlying event is to increase the cross section with respect to the 
parton level, whilst the effect of hadronization is to decrease it. The magnitude of both effects 
decreases asymptotically with increasing \etjet. Below $50~{\rm GeV}$ the hadronization effect dominates, 
leading to an overall decrease of the
cross section with respect to the parton level that is within 10\%. At higher \etjet, the correction is 
driven by the underlying event leading to an increase of at most 5\%. 
A detailed study of this correction is 
outside the scope of this paper, and we do not apply any such corrections to the {\sc MCFM} predictions. 

\begin{figure}[t!!!]
\includegraphics[width=0.49\textwidth]{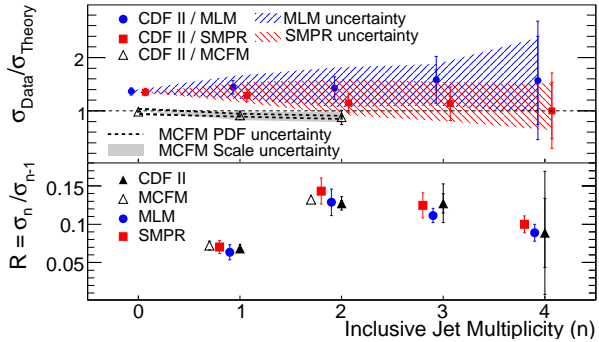}%
\caption{
Top: the ratio of data to theory for the total cross
sections \totalnsigma~ 
as a function of the jet multiplicity $n$. 
Bottom: $\sigma_n/\sigma_{n-1}$ for data, MLM, SMPR and {\sc MCFM}.
Inner (outer) error bars denote the statistical (total) uncertainties
on the measured cross sections.
\label{fig:ratio}}
\end{figure}
\begin{figure}[t!!!]
\includegraphics[width=0.48\textwidth]{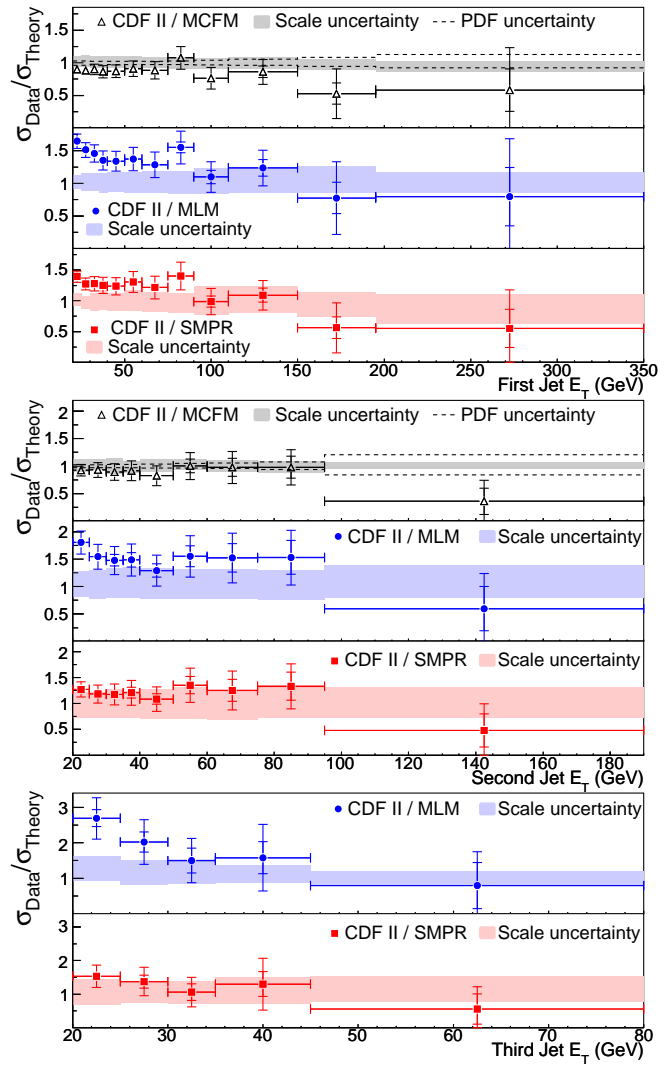}%
\caption{
Ratio of the measured cross sections \diffsigma~to the MLM, SMPR  and {\sc MCFM} predictions 
for $n = 1$ (top), $n = 2$ (middle) and $n = 3$ (bottom). MCFM 
predictions are not available for $n = 3$. Inner (outer) error bars denote the statistical (total) 
uncertainties on the measured cross sections.
\label{fig:diffxsecratio}}	
\end{figure}
%%%%%%%%%%%%%%%%%%
\begin{table}%
\caption{
The measured cross section \diffsigma~ 
for each $E_T^{nth-jet}$-bin ($n=1-4$), with statistical (first) and systematic (second) uncertainties. 
An overall 5.8\% uncertainty on the integrated luminosity is not included.
\label{tab:xsecdiff}}
\begin{ruledtabular}       
\begin{tabular}{lcr}
$E_T^{1st-jet}~$(GeV) ~~ & \multicolumn{2}{r}{~~ $d\sigma/dE_T^{1st-jet}~$(pb/GeV)} \\
\hline
 20$-$25  && 4.46 $\pm $ 0.07 $\pm 0.29$ \\
 25$-$30  && 2.80 $\pm $ 0.06 $\pm 0.21$ \\
 30$-$35  && 1.92 $\pm $ 0.05 $\pm 0.17$ \\
 35$-$40  && 1.31 $\pm $ 0.04 $\pm 0.14$ \\
 40$-$50  && 0.839 $\pm $ 0.023 $\pm 0.093$ \\
 50$-$60  && 0.498 $\pm $ 0.018 $\pm 0.063$ \\
 60$-$75  && 0.259 $\pm $ 0.011 $\pm 0.038$ \\
 75$-$90  && 0.158 $\pm $ 0.008 $\pm 0.024$ \\
 90$-$110 && 0.056 $\pm $ 0.005 $\pm 0.011$ \\
 110$-$150&& 0.0225 $\pm $ 0.0023 $\pm 0.0044$ \\
 150$-$195&& 0.0035 $\pm $ 0.0011 $\pm 0.0023$ \\
 195$-$350&& 0.00039 $\pm $ 0.00022 $\pm 0.00038$ \\
\hline
$E_T^{2nd-jet}~$(GeV) ~~ & \multicolumn{2}{r}{~~ $d\sigma/dE_T^{2nd-jet}~$(pb/GeV)} \\
\hline
 20$-$25  &&  0.874 $\pm $ 0.033 $\pm 0.097$ \\   
 25$-$30  &&  0.483 $\pm $ 0.025 $\pm 0.066$ \\   
 30$-$35  &&  0.286 $\pm $ 0.020 $\pm 0.045$ \\    
 35$-$40  &&  0.190 $\pm $ 0.017 $\pm 0.034$ \\    
 40$-$50  &&  0.095 $\pm $ 0.009 $\pm 0.019$ \\   
 50$-$60  &&  0.057 $\pm $ 0.007 $\pm 0.012$ \\   
 60$-$75  &&  0.0269 $\pm $ 0.0046 $\pm 0.0066$ \\
 75$-$95  &&  0.0107 $\pm $ 0.0022 $\pm 0.0028$ \\
 95$-$190 &&  0.00059 $\pm $ 0.00039 $\pm 0.00049$ \\
\hline
$E_T^{3rd-jet}~$(GeV) ~~ & \multicolumn{2}{r}{~~ $d\sigma/dE_T^{3rd-jet}~$(pb/GeV)} \\
\hline
 20$-$25 &&  0.184 $\pm $ 0.016 $\pm 0.036$ \\	    
 25$-$30 &&  0.087 $\pm $ 0.012 $\pm 0.024$ \\	    
 30$-$35 &&  0.037 $\pm $ 0.008 $\pm 0.013$ \\	    
 35$-$45 &&  0.020 $\pm $ 0.006 $\pm 0.011$ \\
 45$-$80 &&  0.0015 $\pm $ 0.0012 $\pm 0.0013$ \\
\hline
$E_T^{4th-jet}~$(GeV) ~~ & \multicolumn{2}{r}{~~ $d\sigma/dE_T^{4th-jet}~$(pb/GeV)} \\
\hline
20$-$25 && 0.0422 $\pm $ 0.0087 $\pm 0.0079$ \\
25$-$35 && 0.0074 $\pm $ 0.0039 $\pm 0.0036$     \\
\end{tabular}
\end{ruledtabular}
\end{table}

%
%
%
% DISCUSSION OF THEORY COMPARISONS
The upper plot of Fig.~\ref{fig:ratio} shows, as a function of the jet
multiplicity $n$, the ratio of data to theory for the total cross
sections \totalnsigma.
The lower plot shows the ratio $\sigma_n/\sigma_{n-1}$. 
In Fig.~\ref{fig:diffxsecratio} the ratios of the measured differential 
cross sections  \diffsigma~to the predictions 
are shown for $n = 1 - 3$. 
The difference observed in
Fig.~\ref{fig:ratio} between the measured cross sections and SMPR or MLM predictions 
reflects the LO nature of these calculations. 
All the predictions show
good agreement with the data in the cross section ratios $\sigma_n/\sigma_{n-1}$. 
Fig.~\ref{fig:diffxsecratio} shows that 
the variation in the \wnjet~cross section as a function of \etjet~is better reproduced 
by the SMPR prediction than by the MLM. 
A possible explanation is the absence of a tuned underlying event model in the {\sc herwig} component of the MLM prediction. 
We observe  good agreement between the {\sc MCFM} predictions and data in both
total and differential cross section comparisons. 

%%% SUMMARY :
%
In summary, we have used $320$~${\rm pb^{-1}}$ of CDF II data to
measure the differential cross section 
\diffsigmaBR~as a function of the transverse energy of the
$n^{th}$-leading jet, and the total cross section 
\totalnsigmaBR, for $n= 1 - 4$ in a restricted $W \rightarrow e\nu$ decay 
phase space. The cross sections, corrected 
for all detector effects, can be directly compared to the particle level 
predictions of \wjets Monte Carlo generators.

%%% ACKOWLEDGEMENTS :
%
We thank the Fermilab staff and the technical staffs of the participating institutions for their vital contributions.
We also thank J.~Campbell, M.~Mangano, and S.~Mrenna. 
This work was supported by the U.S. Department of Energy and National Science Foundation; the Italian Istituto 
Nazionale di Fisica Nucleare; the Ministry of Education, Culture, Sports, Science and Technology of Japan; 
the Natural Sciences and Engineering Research Council of Canada; the National Science Council of the Republic 
of China; the Swiss National Science Foundation; the A.P. Sloan Foundation; the Bundesministerium 
f\"ur Bildung und Forschung, Germany; the Korean Science and Engineering Foundation and the Korean Research 
Foundation; the Science and Technology Facilities Council and the Royal Society, UK; 
the Institut National de Physique Nucleaire et Physique des Particules/CNRS; 
the Russian Foundation for Basic Research; 
the Comisi\'on Interministerial de Ciencia y Tecnolog\'{\i}a, Spain; 
the European Community's Human Potential Programme; the Slovak R\&D Agency; and the Academy of Finland.
%
%%% REFERENCES %%%
%


\begin{thebibliography}{99}
\bibitem{ref:thesisBDCthesisAM}
  B.~Cooper, Ph.D. thesis, University College London (2006);
  A.~Messina, Ph.D. thesis, Universit\`a di Roma ``La Sapienza'' (2004).
\bibitem{ref:coord}{We use a cylindrical coordinate system 
along the proton direction in which $\theta$~$(\phi)$ is the polar (azimuthal) angle. 
We define~$\eta = -\ln(\tan(\theta/2))$,
$p_{T}= p\sin\theta$, $E_{T}=E\sin\theta$,~\met $=-|\sum_{i}E_{T}^{i}\hat{n}_{i}|$ 
where $\hat{n}_{i}$ is a unit vector in the azimuthal plane that points from the beamline to the $i^{th}$
calorimeter tower, and~$m_T^W=\sqrt{2E_T^{ele}E\!\!\!/_T(1-\rm{cos}\Delta\phi(E_T^{ele},E\!\!\!/_T))}$. 
}
\bibitem{ref:run1Njetrun11jet}
  T.~Affolder {\it et al.}  (CDF Collaboration),
  %``Test of enhanced leading order QCD in $W$ boson plus jets events from  1.8
  %TeV $\bar{p}p$ collisions,''
  Phys.\ Rev.\  D {\bf 63}, 072003 (2001);
  F.~Abe {\it et al.}  (CDF Collaboration),
  %``Measurement of the $\sigma(W + \ge 1 {\rm jet})/\sigma(W)$ cross section
  %ratio from $\bar{p}p$ collisions at $\sqrt{s} = 1.8$ TeV,''
  Phys.\ Rev.\ Lett.\  {\bf 81}, 1367 (1998).
\bibitem{ref:d0zjets}
  V.~M.~Abazov {\it et al.}  (D0 Collaboration),
  %``Measurement of the ratios of the Z/gamma* + >= $n$ jet production cross
  %sections to the total inclusive Z/gamma* cross section in $p \bar{p}$
  %collisions at $\sqrt{s}$ = 1.96-TeV,''
  accepted by Phys.\ Lett.\ B, arXiv:hep-ex/0608052 (2006).
\bibitem{ref:particlelevel}
  A.~Abulencia {\it et al.} (CDF Collaboration),
  Phys.\ Rev.\  D {\bf 74}, 071103(R) (2006).
\bibitem{ref:CDFb}
  D.~Acosta {\it et al.}  (CDF Collaboration),
  %``Measurement of the $J/\psi$ meson and $b-$hadron production cross sections
  %in $p\bar{p}$ collisions at $\sqrt{s} = 1960$ GeV,''
  Phys.\ Rev.\  D {\bf 71}, 032001 (2005).
\bibitem{ref:tightsel}
  A.~Abulencia {\it et al.}  (CDF Collaboration),
  %``Measurements of inclusive $W$ and $Z$ cross sections in $p\bar{p}$
  %collisions  at $\sqrt{s}=1.96$ TeV,''
  J.\ Phys.\ G: Nucl.\ Part.\ Phys. {\bf 34}, 2457 (2007).
\bibitem{ref:corrMet}
  A.~Abulencia {\it et al.}  (CDF Collaboration),
  %``Measurement of the $t \bar{t}$ Production Cross Section in $p \bar{p}$
  %collisions at $\sqrt{s}$ = 1.96-TeV using Lepton + Jets Events with Jet
  %Probability $b^-$ tagging,''
  Phys.\ Rev.\  D {\bf 74}, 072006 (2006).
\bibitem{ref:jetalg}
  F.~Abe {\it et al.}  (CDF Collaboration),
  %``The Topology of three jet events in $\bar{p}p$ collisions at $\sqrt{s} =
  %1.8$ TeV,''
  Phys.\ Rev.\  D {\bf 45}, 1448 (1992).
\bibitem{ref:jetcorr}
  A.~Bhatti {\it et al.},
  %``Determination of the jet energy scale at the Collider Detector at
  %Fermilab,''
  Nucl.\ Instrum.\ Methods\  A {\bf 566}, 375 (2006).
\bibitem{ref:Wlikexsec}
  W.-M.~Yao {\it et al.},
  %``Review of particle physics,''
  J.\ Phys.\ G: Nucl.\ Part.\ Phys. {\bf 33}, 1 (2006).
\bibitem{ref:dibosonxsectopxsec}
  M.~S.~Neubauer  (CDF and D0 Collaborations),
  %``Diboson physics at the Tevatron,''
  arXiv:hep-ex/0605066 (2006);
  A.~Abulencia {\it et al.}  (CDF Collaboration),
  %``Measurement of the tanti-t Production Cross Section in $p$ anti-ptnipbar
  %Collisions at $\sqrt{s}$ = 1.96-TeV,''
  Phys.\ Rev.\ Lett.\  {\bf 97}, 082004 (2006).
\bibitem{ref:alpgen}
  M.~L.~Mangano {\it et al.},
  %``ALPGEN, a generator for hard multiparton processes in hadronic
  %collisions,''
  J. High Energy Phys. {\bf 0307}, 001 (2003).
\bibitem{ref:pythia}
  T.~Sjostrand, S.~Mrenna, and P.~Skands,
  %``PYTHIA 6.4 physics and manual,''
  J. High Energy Phys. {\bf 0605}, 026 (2006).
\bibitem{ref:TUNEA}
  D.~Acosta {\it et al.}  (CDF Collaboration),
  %``The underlying event in hard interactions at the Tevatron $\bar{p}p$
  %collider,''
  Phys.\ Rev.\  D {\bf 70}, 072002 (2004).
\bibitem{ref:herwig}
  G.~Corcella {\it et al.},
  %``HERWIG 6: An event generator for hadron emission reactions with
  %interfering gluons (including supersymmetric processes),''
  J. High Energy Phys. {\bf 0101}, 010 (2001).
\bibitem{ref:madgraph}
  F.~Maltoni and T.~Stelzer,
  %``MadEvent: Automatic event generation with MadGraph,''
  J. High Energy Phys. {\bf 0302}, 027 (2003).
\bibitem{ref:gleisberg2003lavesson2005papadopoulos2005}
  T.~Gleisberg {\it et al.},
  %``SHERPA 1.alpha, a proof-of-concept version''
  J. High Energy Phys. {\bf 0402}, 056 (2004);
  N.~Lavesson and L.~Lonnblad,
  %``W + jets matrix elements and the dipole cascade,''
  J. High Energy Phys. {\bf 0507}, 054 (2005);
  C.~G.~Papadopoulos and M.~Worek,        
  %``Multi-parton Cross Sections at Hadron Colliders''
  Eur.\ Phys.\ J.\  C {\bf 50}, 843 (2007).
\bibitem{ref:CKKWkrauss2002}
  S.~Catani {\it et al.},
  %``QCD matrix elements + parton showers,''
  J. High Energy Phys. {\bf 0111}, 063 (2001);
  F.~Krauss,
  %``Matrix Elements and Parton Showers in Hadronic Interaction''
  J. High Energy Phys. {\bf 0208}, 015 (2002).
\bibitem{ref:lonnblad2002}
  L.~Lonnblad,            
  %``Correcting the colour-dipole cascade model with fixed order matrix
  %elements,''
  J. High Energy Phys. {\bf 0205}, 046 (2002).
\bibitem{ref:mangano2006}
  M.~L.~Mangano {\it et al.},          
  %``Matching matrix elements and shower evolution for top-production in hadronic collisions''
  J. High Energy Phys. {\bf 0701}, 013 (2007).
\bibitem{ref:mrenna}
  S.~Mrenna and P.~Richardson,
  %``Matching matrix elements and parton showers with HERWIG and PYTHIA,''
  J. High Energy Phys. {\bf 0405}, 040 (2004).
\bibitem{ref:alwall2007}
  J.~Alwall {\it et al.},
  %``Comparative study of various algorithms for the merging of parton   showers
  %and matrix elements in hadronic collisions''
  arXiv:0706.2569 [hep-ph] (2007).
\bibitem{ref:pumplin2002}
  J.~Pumplin {\it et al.},
  J. High Energy Phys. {\bf 0207}, 012 (2002).
\bibitem{ref:lai2000}
  H.~L.~Lai {\it et al.},
  %``Global {QCD} analysis of parton structure of the nucleon: CTEQ5 parton
  %distributions,''
  Eur.\ Phys.\ J.\  C {\bf 12}, 375 (2000).
\bibitem{ref:MCNLO}
  J.~Campbell and R.~K.~Ellis,
  %``Next-to-leading order corrections to W + 2jet and Z + 2jet production  at
  %hadron colliders,''
  Phys.\ Rev.\  D {\bf 65} 113007 (2002).
\bibitem{ref:Ellis1993tq}
  S.~D.~Ellis and D.~E.~Soper,
    Phys.\ Rev.\  D {\bf 48} 3160 (1993).
\bibitem{ref:campbell2006}
  J.~M.~Campbell, J.~W.~Huston, and W.~J.~Stirling,
  %``Hard interactions of quarks and gluons: A primer for LHC physics,''
  Rep.\ Prog.\ Phys.\  {\bf 70}, 89 (2007).
\end{thebibliography}
\end{document}